\documentclass[epj]{svjour}
\usepackage{graphics}
\begin{document}
\title{Understanding one-dimensional topological Kondo insulator: Poor man's non-uniform antiferromagnetic mean-field theory versus quantum Monte Carlo simulation}
\author{Yin Zhong\inst{1}\thanks{\emph{Present address:} zhongy05@hotmail.com}}
\institute{Center for Interdisciplinary Studies $\&$ Key Laboratory for
Magnetism and Magnetic Materials of the MoE, Lanzhou University, Lanzhou 730000}
\date{Received: date / Revised version: date}
\abstract{Topological Kondo insulator (TKI) is an essential example of interacting topological insulator, where electron's correlation effect plays a key role. However, most of our understanding on this timely issue comes from numerical simulations, (particularly in one-spatial dimension) which exactly includes correlation effect but is black box for extracting underlying physics. In this work, we use a non-uniform antiferromagnetic mean-field (nAFM) theory to understand the underlying physics in a TKI model, the $1D$ $p-$wave periodic Anderson model ($p$-PAM). Comparing with numerically exact quantum Monte Carlo simulation, we find that nAFM theory is an excellent approximation for ground-state properties when onsite Hubbard interaction is weak. This emphasizes the dominating antiferromagnetic correlation in this system and local antiferromagnetic picture captures the qualitative nature of interacting many-body ground state. Adding extra conduction electron band to $p$-PAM leads to a quantum phase transition from Haldane phase into topological trivial phase. We believe these results may be helpful for understanding novel physics in interacting TKI materials such as SmB$_{6}$ and other related compounds.
} 
\maketitle
\section{Introduction}\label{intr}
In recent years, topological states of matter has been the mainstream of condensed matter physics after the discovery of $2D$ quantum spin Hall effect, $3D$ topological insulator and topological semimetal.\cite{Hasan2010,Qi2011,Armitage2018}
The electronic structures of these real-life topological materials have been successfully described with topological band theory,\cite{Bansil2016} which is based on non-interacting single-electron picture.

In contrast, strongly interacting topological materials like topological Kondo
insulator (TKI) candidate SmB$_{6}$,\cite{Dzero2016,Dzero2010} are still poorly understood due to intrinsic electron correlation.\cite{Li2014,Tan2015}
Much efforts have been made to understand the anomaly observed in SmB$_{6}$ and many novel ideas emerge like surface Kondo breakdown, Majorana Fermi sea, failed superconductor, fractionalized Fermi liquid and composite exciton.\cite{Alexandrov2015,Erten2016,Baskaran2015,Erten2017,Thomson2016,Chowdhury2017,Sodemann2017}

But, due to lack of controllable theory to treat electron's correlation effect, reliability or relevance of these theories are still unknown. Fortunately, exact numerical simulations including static and dynamic electron correlation can provide benchmark for various approximations and may clarify the nature of these novel theories.

Very recently, we have taken a step in this direction by performing a zero-temperature quantum Monte Carlo (QMC) simulation on one-dimensional $p$-wave periodic Anderson model ($p$-PAM).\cite{Zhong2017} The $1D$ $p$-PAM is a simplified model to understand electron correlation effect in TKI.
We find that the ground-state is the Haldane phase though the non-interacting limit corresponds to a $Z_{2}$ topological insulator. Furthermore, these results are verified by an independent density matrix renormalization group study.\cite{Lisandrini2017} In addition, we have also studied the finite temperature physics of $p$-PAM by finite-$T$ QMC simulation,\cite{Zhong2017b} and its nonequilibrium dynamics has been calculated in Ref.\cite{Hagymasi2019}.

Given these inspiring numerical results, it is tempting to extract intuitive physics from those black boxes. For this purpose, in this work, we use a non-uniform antiferromagnetic mean-field (nAFM) theory to understand the underlying physics in $1D$ $p$-PAM. We find that nAFM theory is an excellent approximation for ground state when Hubbard interaction is weak. Specifically, the key physical quantities like site-resolved magnetization, double occupation number of $f$-electron and $c-f$ hybridization strength are all comparable to QMC. Beside the original model, we add an extra conduction electron band to $p$-PAM, which provides a quantum phase transition from Haldane phase into topological trivial phase.
It is believed that these results may be helpful for understanding novel physics in interacting topological Kondo insulator like SmB$_{6}$ and other related materials.
\begin{figure}
\resizebox{0.55\textwidth}{!}{%
  \includegraphics{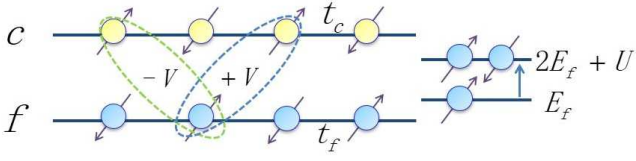}%
}
\caption{$1D$ $p$-wave periodic Anderson model describes a $p$-wave-like hybridization $\pm V$ between local f-electron orbital (blue) and its neighboring conducting charge carrier (yellow).
The singly occupied local f-electron has energy $E_{f}$ while double occupation has extra Coloumb energy $U$. The conduction (local) electron hops $t_{c}$ ($-t_{f}$) between nearest-neighbor sites.}
\label{fig:PAM}
\end{figure}
\section{Model and Mean-field approximation}\label{sec2}
The $1D$ $p$-PAM has the following Hamiltonian:\cite{Zhong2017}
\begin{eqnarray}
H&&=\sum_{j\sigma}[t_{c}\hat{c}_{j\sigma}^{\dag}\hat{c}_{j+1\sigma}-t_{f}\hat{f}_{j\sigma}^{\dag}\hat{f}_{j+1\sigma}+\mathrm{H.c.}]\nonumber \\
&&+\frac{V}{2}\sum_{j\sigma}[(\hat{c}_{j+1\sigma}^{\dag}-\hat{c}_{j-1\sigma}^{\dag})\hat{f}_{j\sigma}+\hat{f}_{j\sigma}^{\dag}(\hat{c}_{j+1\sigma}-\hat{c}_{j-1\sigma})]\nonumber \\
&&+E_{f}\sum_{j\sigma}\hat{f}_{j\sigma}^{\dag}\hat{f}_{j\sigma}+U\sum_{j}\hat{f}_{j\uparrow}^{\dag}\hat{f}_{j\uparrow}\hat{f}_{j\downarrow}^{\dag}\hat{f}_{j\downarrow}\label{eq1}.
\end{eqnarray}
Here, $t_{c}$ and $t_{f}$ are the nearest-neighbor-hopping strengths along the one-dimensional lattice for conduction and $f$-electron, respectively. $\hat{c}_{j\sigma}$ ($\hat{f}_{j\sigma}$) is the fermion annihilation operator for conduction electron ($f$-electron).  The $f$-electron has single-particle energy level $E_{f}$ and the on-site Hubbard interaction $U$.
To give a $1D$ TKI, a $p$-wave-like hybridization between conduction and $f$-electron (the $V$ term) is introduced. (See also Fig.~\ref{fig:PAM})

Now, we consider mean-field decoupling of $1D$ $p$-PAM. The merit of mean-field treatment is to decouple interaction term into quadratic form, for the Hubbard interaction encountered here, one can rewrite it as follows
\begin{eqnarray}
\hat{f}_{j\uparrow}^{\dag}\hat{f}_{j\uparrow}\hat{f}_{j\downarrow}^{\dag}\hat{f}_{j\downarrow}&&=-\frac{1}{4}(\hat{f}_{j\uparrow}^{\dag}\hat{f}_{j\uparrow}-\hat{f}_{j\downarrow}^{\dag}\hat{f}_{j\downarrow})^{2}+\frac{1}{4}(\hat{f}_{j\uparrow}^{\dag}\hat{f}_{j\uparrow}+\hat{f}_{j\downarrow}^{\dag}\hat{f}_{j\downarrow})\nonumber\\
&&\simeq-\frac{1}{4}\left[2m_{j}^{f}(\hat{f}_{j\uparrow}^{\dag}\hat{f}_{j\uparrow}-\hat{f}_{j\downarrow}^{\dag}\hat{f}_{j\downarrow})-(m_{j}^{f})^{2}\right]\nonumber\\
&&+\frac{1}{4}\left[2n_{j}^{f}(\hat{f}_{j\uparrow}^{\dag}\hat{f}_{j\uparrow}+\hat{f}_{j\downarrow}^{\dag}\hat{f}_{j\downarrow})-(n_{j}^{f})^{2}\right]\nonumber
\end{eqnarray}
where we have defined the magnetic (density) order parameter $m_{j}^{f}$ and charge (density) order parameter $n_{j}^{f}$ via
\begin{eqnarray}
m_{j}^{f}=\langle\hat{f}_{j\uparrow}^{\dag}\hat{f}_{j\uparrow}-\hat{f}_{j\downarrow}^{\dag}\hat{f}_{j\downarrow}\rangle, ~~n_{j}^{f}=\langle\hat{f}_{j\uparrow}^{\dag}\hat{f}_{j\uparrow}+\hat{f}_{j\downarrow}^{\dag}\hat{f}_{j\downarrow}\rangle.\nonumber
\end{eqnarray}
Therefore, the Hubbard interaction can be approximated as
\begin{eqnarray}
\hat{f}_{j\uparrow}^{\dag}\hat{f}_{j\uparrow}\hat{f}_{j\downarrow}^{\dag}\hat{f}_{j\downarrow}\simeq\sum_{\sigma}\left(-\frac{m_{j}^{f}}{2}\sigma+\frac{n_{j}^{f}}{2}\right)\hat{f}_{j\sigma}^{\dag}\hat{f}_{j\sigma}+\frac{(m_{j}^{f})^{2}-(n_{j}^{f})^{2}}{4}\nonumber
\end{eqnarray}
and the $p$-PAM under mean-field decoupling is found to be
\begin{eqnarray}
\hat{H}_{MF}&&=\sum_{j\sigma}[t_{c}\hat{c}_{j\sigma}^{\dag}\hat{c}_{j+1\sigma}-t_{f}\hat{f}_{j\sigma}^{\dag}\hat{f}_{j+1\sigma}+\mathrm{H.c.}]\nonumber\\
&&+\frac{V}{2}\sum_{j\sigma}[(\hat{c}_{j+1\sigma}^{\dag}-\hat{c}_{j-1\sigma}^{\dag})\hat{f}_{j\sigma}+\hat{f}_{j\sigma}^{\dag}(\hat{c}_{j+1\sigma}-\hat{c}_{j-1\sigma})]\nonumber \\
&&+\sum_{j\sigma}\left(E_{f}-U\frac{m_{j}^{f}}{2}\sigma+U\frac{n_{j}^{f}}{2}\right)\hat{f}_{j\sigma}^{\dag}\hat{f}_{j\sigma}\nonumber\\
&&+U\sum_{j}\frac{(m_{j}^{f})^{2}-(n_{j}^{f})^{2}}{4}.\label{eq2}
\end{eqnarray}
To proceed, we recall that in our previous QMC simulations on half-filled symmetric $p$-PAM, the leading correlation is antiferromagnetic,\cite{Zhong2017} so the magnetic order parameter can be embodied as the antiferromagnetic order parameter $m_{j}^{f}=(-1)^{j}m_{j}^{f}$. Note that both $m_{j}^{f}$ and $n_{j}^{f}$ are site-dependent, thus it permit non-uniform distribution of antiferromagnetic order and possible charge order.

Because we are interested in topological states in $p$-PAM (e.g. Haldane phase), the open boundary condition (OBC) will be used in this work, which is able to detect the edge local moment in Haldane phase.\cite{Mezio2015,Hagymasi2016} Therefore, the mean-field Hamiltonian should be solved in a finite lattice system with OBC.

Since the system works in real-space, it is not able to write down an explicit mean-field equations. But, we can solve this mean-field theory as follows. Firstly, we just guess a trial solution of $m_{j}^{f},n_{j}^{f}$, then inserting it into mean-field Hamiltonian Eq.\ref{eq2}.
By diagonalizing Hamiltonian Eq.\ref{eq2} in site basis, we can obtain all single-particle orbits. Then, by using these single-particle orbits, one can construct the ground-state wavefunction (Slater determinant) for given electron and spin density. Next, all expectation values like order parameter $m_{j}^{f},n_{j}^{f}$ are readily to find by their definition. Finally, we use these new $m_{j}^{f},n_{j}^{f}$ to replace old ones to form a iteration loop. When convergence is reached, the mean-field theory is solved and the ultimate $m_{j}^{f},n_{j}^{f}$ are the required solutions.

\section{Mean-field solution versus QMC simulation}
To meet with QMC simulation, we here focus on the half-filled symmetric $p$-PAM.\cite{Zhong2017} This means the chemical potential for conduction and f-electron are setting to zero and f-electron energy level is fixed to be $E_{f}=-U/2$.
\begin{figure}
\resizebox{0.58\textwidth}{!}{%
  \includegraphics{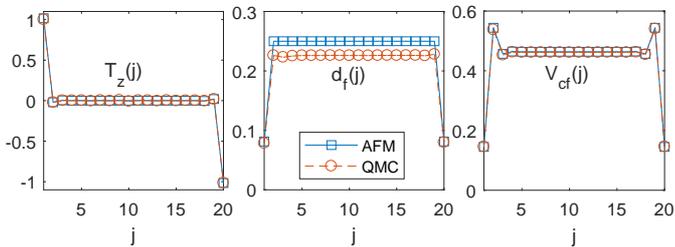}%
}
\caption{Antiferromagnetic mean-field (AFM) solution versus QMC simulation for site-resolved magnetization $T_{z}(j)$, double occupation number of f-electron $d_{f}(j)$ and $c-f$ hybridization $V_{cf}(j)$. Parameters are $t_{c}=1,t_{f}=\pi/10,V=1,U=0.5,E_{f}=-0.25$.}
\label{fig:figure22}
\end{figure}

In Fig.\ref{fig:figure22}, we have given an example of the non-uniform AFM solution. Here, we examine the following physical quantities: site-resolved magnetization
$T_{z}(j)=\sum_{\sigma}\sigma\langle\hat{f}_{j\sigma}^{\dag}\hat{f}_{j\sigma}+\hat{c}_{j\sigma}^{\dag}\hat{c}_{j\sigma}\rangle,$
double occupation number of f-electron
$d_{f}(j)=\langle \hat{f}_{j\uparrow}^{\dag}\hat{f}_{j\uparrow}\hat{f}_{j\downarrow}^{\dag}\hat{f}_{j\downarrow}\rangle$
and effective $c-f$ hybridization $V_{cf}(j)=-\frac{1}{2}\sum_{\sigma}\langle \hat{f}_{j\sigma}^{\dag}\hat{c}_{j+1\sigma}-\hat{f}_{j\sigma}^{\dag}\hat{c}_{j-1\sigma}\rangle.$
A $20$-site chain is considered in the present case and other system parameters are $t_{c}=1,t_{f}=\pi/10,V=1,U=0.5,E_{f}=-0.25$. Moreover, as comparison, the $T=0$ QMC simulation uses an imaginary-time evolving time $\beta=50$, imaginary-time interval $\Delta\tau=0.1$. We have tested
a longer chain and larger $\beta$, and it does not lead to any sensible changes.

From Fig.\ref{fig:figure22}, the agreement between nAFM solution and QMC is excellent, particularly for $T_{z}(j)$ and $V_{cf}(j)$. This suggests that the edge local moment/magnetization can be considered as the magnetic order of edge electrons while the bulk electrons have no such magnetic order and no sensible magnetization are observed. However, since charge fluctuation is underestimated due to the crude decoupling of Hubbard interaction, $d_{f}(j)$ is overestimated in the mean-field theory. To improve, slave-particle approach like $Z_{2}$ slave-spin mean-field theory should be helpful.\cite{Huber2012}
\begin{figure}
\resizebox{0.4\textwidth}{!}{%
  \includegraphics{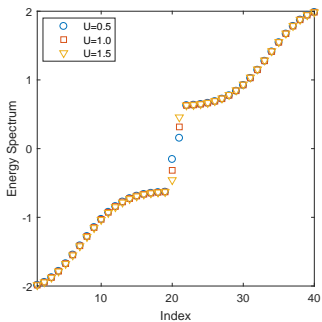}%
}
\caption{The single-particle energy spectrum for different interaction $U=0.5,1.0$ and $1.5$. Other parameters are the same with Fig.\ref{fig:figure22}.}
\label{fig:eigenenergy}
\end{figure}

Furthermore, because the system is approximated by the mean-field Hamiltonian Eq.\ref{eq2}, all single-particle eigen-energy can be obtained as shown in in Fig.\ref{fig:eigenenergy}. Here, the bulk electron band and the edge state around zero energy are shown. It is known that when the Hubbard interaction is turned off ($U=0$), there exist two degenerated zero energy modes, (for each spin flavor) which is the localized edge mode located around boundary.\cite{Zhong2017} As seen in Fig.\ref{fig:eigenenergy}, when reintroducing interaction, the bulk modes are not changed. In contrast, the energy of edge mode is deviated from zero and shifts toward bulk band mode. This is due to the formation of antiferromagnetic order around edges. When interaction is large enough, these modes are expected to immerse into the bulk band.

\begin{figure}
\resizebox{0.5\textwidth}{!}{%
  \includegraphics{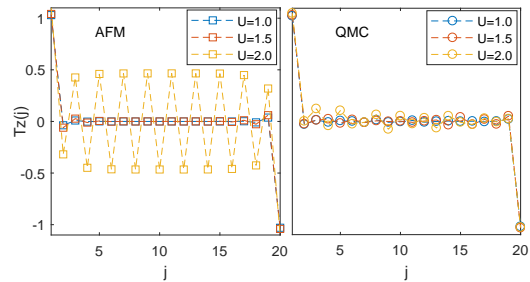}%
}
\caption{The site-resolved magnetization $T_{z}(j)$ in AFM theory (left) and QMC simulation (right) for different Hubbard interaction $U$. Other parameters are the same with Fig.\ref{fig:figure22}.}
\label{fig:figurecom}
\end{figure}

However, as can be seen in Fig.\ref{fig:figurecom}, when interaction is further enhanced, the nAFM theory predicts that the bulk of the system turns out to have antiferromagnetic order, which should be prohibited by wild quantum fluctuation in $1D$ and is contrast to the results of QMC.
Therefore, the nAFM theory is reliable at weak coupling ($U\leq t_{c}$) and it gives rise to wrong long-ranged antiferromagnetic ordered states when interaction is larger.

A careful reader may notice that the mean-field approach like nAFM should provide even better results for the $3D$ TKI models, because of the weakened quantum fluctuations. But we have to emphasize that although the mean-field treatment is more reliable in $3D$, the TKI state in $3D$ is a paramagnetic insulator and its edge/surface state is also paramagnetic. Thus, if we extend our antiferromagnetic mean-field to $3D$ case, the predicted edge magnetization will lead to magnetic order, which is in contrast to experiments and theoretical calculation.\cite{Dzero2016}
\section{Model with extra conduction electron band}
\begin{figure}
\resizebox{0.55\textwidth}{!}{%
  \includegraphics{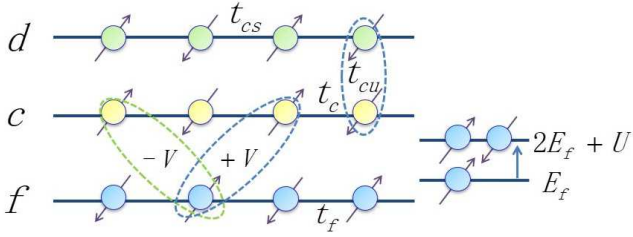}%
}
\caption{$1D$ $p$-PAM couples with an extra conduction electron band (green). The hoping strength of this extra conduction electron band is $t_{cs}$ and its coupling to original band is $t_{cu}$.}
\label{fig:PAMex}
\end{figure}

For $1D$ $p$-PAM, its ground-state is the well-established Haldane phase.\cite{Zhong2017,Lisandrini2017}
Since it is a (symmetry-protected) topological state, the Haldane phase itself is robust against weak interaction and perturbation.
So, it is interesting to see whether there exists a quantum phase transition from Haldane phase to other non-trivial or trivial state of matter
in the $1D$ $p$-PAM. Here, we consider a simple realization, where the $1D$ $p$-PAM couples with an extra conduction electron band, whose model Hamiltonian reads as follows:(See also Fig.~\ref{fig:PAMex})
\begin{eqnarray}
H&&=\sum_{j\sigma}[t_{c}\hat{c}_{j\sigma}^{\dag}\hat{c}_{j+1\sigma}-t_{f}\hat{f}_{j\sigma}^{\dag}\hat{f}_{j+1\sigma}+\mathrm{H.c.}]\nonumber \\
&&+\frac{V}{2}\sum_{j\sigma}[(\hat{c}_{j+1\sigma}^{\dag}-\hat{c}_{j-1\sigma}^{\dag})\hat{f}_{j\sigma}+\hat{f}_{j\sigma}^{\dag}(\hat{c}_{j+1\sigma}-\hat{c}_{j-1\sigma})]\nonumber \\
&&+E_{f}\sum_{j\sigma}\hat{f}_{j\sigma}^{\dag}\hat{f}_{j\sigma}+U\sum_{j}\hat{f}_{j\uparrow}^{\dag}\hat{f}_{j\uparrow}\hat{f}_{j\downarrow}^{\dag}\hat{f}_{j\downarrow}\nonumber\\
&&+t_{cs}\sum_{j\sigma}[\hat{d}_{j\sigma}^{\dag}\hat{d}_{j+1\sigma}+\hat{d}_{j+1\sigma}^{\dag}\hat{d}_{j\sigma}]\nonumber\\
&&+t_{cu}\sum_{j\sigma}[\hat{d}_{j\sigma}^{\dag}\hat{c}_{j\sigma}+\hat{c}_{j\sigma}^{\dag}\hat{d}_{j\sigma}]
\label{eq3}.
\end{eqnarray}
Here, $\hat{d}_{j\sigma}$ is the annihilation operator for the second conduction electron band. The hopping energy for this band is $t_{cs}$ and its coupling to the first conduction electron band is onsite with strength $t_{cu}$.

\begin{figure}
\resizebox{0.4\textwidth}{!}{%
  \includegraphics{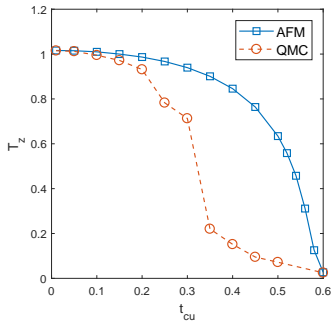}%
}
\caption{The edge magnetization $T_{z}$ calculated by nAFM theory for different conduction electron coupling $t_{cu}$. As comparison, data from QMC is also shown.}
\label{fig:Tz}
\end{figure}

Because the non-zero edge magnetization is an essential feature of Haldane phase, we have examined this quantity for the above model in terms of nAFM theory. In Fig.\ref{fig:Tz}, the edge magnetization $T_{z}$ is shown for different conduction electron coupling $t_{cu}$. (We consider weak coupling case with $U=0.5$, where nAFM theory can give reliable results.)

It is clear that with the increasing of $t_{cu}$, the edge magnetization gradually decreases and finally vanishes when $t_{cu}$ is large. The global
behaviors from nAFM is similar to QMC though the latter predicts a smaller critical $t_{cu}$. Actually, the observed evolution can be understood as the immersing of edge magnetization into the extra conduction band. When $t_{cu}=0$, there are edge modes with definite spin orientation, which leads to edge magnetization. If $t_{cu}$ is finite, the coupling with extra conduction electron band delocalizes the edge mode since these conduction electrons are rather itinerant and has no topological protection like the $p$-wave hybridization. Then, delocalization will be enhanced by increasing $t_{cu}$ and at last only a fully itinerant mode can be found.

\begin{figure}
\resizebox{0.5\textwidth}{!}{%
  \includegraphics{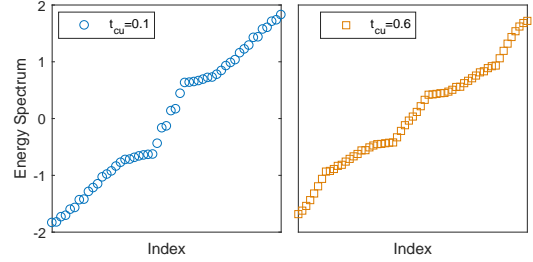}%
}
\caption{The single-particle energy spectrum at small and large $t_{cu}$.}
\label{fig:es}
\end{figure}

At the same time, by inspecting the single-particle energy spectrum at small and large $t_{cu}$, (see Fig.\ref{fig:es}) we find that the system at small $t_{cu}$ is still the insulating Haldane phase while the large $t_{cu}$ case corresponds to a metallic state.
Therefore, we conclude that when increasing the coupling $t_{cu}$, there exists a quantum phase transition from insulating Haldane phase to metallic trivial state.

\section{Conclusion and direction for future work}
In summary, by comparing with numerically exact QMC simulation, the non-uniform AFM theory provides a good description for ground-state properties in typical $1D$ TKI model (the $p$-PAM) when onsite Hubbard interaction is large. This emphasizes the dominating anti-ferromagnetic correlation in this system and local antiferromagnetic picture captures the qualitative nature of interacting many-body ground state. Furthermore, when extra conduction electron band is added, the non-uniform AFM treatment predicts a quantum phase transition from (topological) Haldane state into trivial metallic state. We think these findings should be helpful for understanding novel physics in interacting topological Kondo insulator such as SmB$_{6}$ and other related materials.

\section{Acknowledgments}
This research was supported in part by NSFC under Grant No.~$11704166$, No.~$11834005$ and the Fundamental Research Funds for the Central Universities.

\section{Contribution statement}
Y. Zhong suggested the issue and carried out the calculation. All of authors wrote and revised this article.


\begin{thebibliography}{}
\bibitem{Hasan2010}
M. Z. Hasan and C. L. Kane,  Rev. Mod. Phys. \textbf{82}, 3045 (2010).

\bibitem{Qi2011}
X.-L. Qi and S.-C. Zhang,  Rev. Mod. Phys. \textbf{83}, 1057 (2011).

\bibitem{Armitage2018}
N. P. Armitage, E. J. Mele and A. Vishwanath, Rev. Mod. Phys. \textbf{90}, 015001 (2018).

\bibitem{Bansil2016}
A. Bansil, H. Lin and T. Das,  Rev. Mod. Phys. \textbf{88}, 021004 (2016).

\bibitem{Dzero2016}
M. Dzero, J. Xia, V. Galitski and P. Coleman, Annu. Rev. Condens. Matter Phys. \textbf{7}, 249 (2016).

\bibitem{Dzero2010}
M. Dzero, K. Sun, V. Galitski and P. Coleman, Phys. Rev. Lett. \textbf{104}, 106408 (2010).

\bibitem{Li2014}
G. Li et al., Science \textbf{346}, 1208 (2014).

\bibitem{Tan2015}
B. S. Tan et al., Science \textbf{349}, 287 (2015).

\bibitem{Alexandrov2015}
V. Alexandrov, P. Coleman and O. Erten, Phys. Rev. Lett. \textbf{114}, 177202 (2015).

\bibitem{Erten2016}
O. Erten, P. Ghaemi and P. Coleman, Phys. Rev. Lett. \textbf{116}, 046403 (2016).

\bibitem{Baskaran2015}
G. Baskaran, arXiv:1507.03477.

\bibitem{Erten2017}
O. Erten, P.-Y. Chang, P. Coleman and A. M. Tsvelik, Phys. Rev. Lett. \textbf{119}, 057603 (2017).

\bibitem{Thomson2016}
A. Thomson and S. Sachdev, Phys. Rev. B \textbf{93}, 125103 (2016).

\bibitem{Chowdhury2017}
D. Chowdhury, I. Sodemann and T. Senthil, Nat. Commun. \textbf{9}, 1766 (2018).

\bibitem{Sodemann2017}
I. Sodemann, D. Chowdhury and T. Senthil, Phys. Rev. B \textbf{97}, 045152 (2018).

\bibitem{Zhong2017}
Y. Zhong, Y. Liu and H.-G. Luo, Eur. Phys. J. B \textbf{90}, 147 (2017) (2017).

\bibitem{Lisandrini2017}
F. T. Lisandrini, A. M. Lobos, A. O. Dobry and C. J. Gazza, Phys. Rev. B \textbf{96}, 075124 (2017).

\bibitem{Zhong2017b}
Y. Zhong, Y. Liu, Q. Wang, K. Liu, H.-F. Song and H.-G. Luo, Front. Phys. \textbf{14}, 23602 (2019).

\bibitem{Hagymasi2019}
I. Hagym\'{a}si, C. Hubig and U. Schollw\"{o}ck, Phys. Rev. B \textbf{99}, 075145 (2019).

\bibitem{Mezio2015}
A. Mezio, A. M. Lobos, A. O. Dobry and C. J. Gazza, Phys. Rev.B \textbf{92}, 205128 (2015).

\bibitem{Hagymasi2016}
I. Hagymasi and O. Legeza, Phys. Rev. B \textbf{93}, 165104 (2016).

\bibitem{Huber2012}
A. R\"{u}egg, S. D. Huber and M. Sigrist, Phys. Rev. B \textbf{81}, 155118 (2012).


\end{thebibliography}
\end{document}